\documentclass[prl,twocolumn,preprintnumbers,amsmath,amssymb,nopacs,nofootinbib,floatfix]{revtex4} 
\usepackage{graphicx,bm}

\makeatletter
\def\graphicscale{\twocolumn@sw{0.3}{0.4}}
\def\graphicthreescale{\twocolumn@sw{0.3}{0.4}}

\begin{document}

\title{Anomalous finite-size scaling at thermal first-order
  transitions\\
 in systems with disordered boundary conditions}

\author{Haralambos Panagopoulos,$^1$ Andrea Pelissetto,$^2$ and Ettore Vicari$^3$} 

\address{$^1$ Department of Physics, University of Cyprus,
        Lefkosia, CY-1678, Cyprus} 

\address{$^2$ Dipartimento di Fisica dell'Universit\`a di Roma ``La
  Sapienza'' and INFN, Sezione di Roma I, I-00185 Roma, Italy}

\address{$^3$ Dipartimento di Fisica dell'Universit\`a di Pisa
        and INFN, Largo Pontecorvo 3, I-56127 Pisa, Italy}
\date{\today}

\begin{abstract}

We investigate the equilibrium and off-equilibrium behaviors of
systems at thermal first-order transitions (FOTs) when the boundary
conditions favor one of the two phases.  As a theoretical laboratory
we consider the two-dimensional Potts model.  We show that an
anomalous finite-size scaling emerges in systems with open boundary
conditions favoring the disordered phase, associated with a mixed
regime where the two phases are spatially separated.  Correspondingly,
if the system is slowly heated across the transition, the
characteristic times of the off-equilibrium dynamics scale with a
power of the size.  We argue that these features generally apply to
systems at FOTs, when boundary conditions favor one of the two
phases. In particular, they should be relevant for the experimental
search of FOTs of the quark-gluon plasma in heavy-ion collisions.

\end{abstract}

\maketitle


Understanding finite-size effects at phase transitions is of great
phenomenological importance, because it allows us to interpret
correctly experiments and numerical investigations of finite-size
systems close to the transition point, where thermodynamic quantities
develop singularities in the infinite-volume limit.  At continuous
transitions finite-size scaling (FSS) ~\cite{Barber-83,Privman-90} is
characterized by universal power laws, with critical exponents that
are independent of the geometry and of the boundary conditions, the
latter affecting only FSS functions and amplitudes. In this respect
first-order transitions (FOTs) are more complicated. Most theoretical
studies, see, e.g., Refs.~\cite{NN-75,FB-82,CLB-86,BK-90,VRSB-93},
have considered cubic $L^d$ systems with periodic boundary conditions
(PBC), showing that finite-size effects are generally characterized by
power laws related to the space dimension of the system; for instance,
the correlation-length exponent is $\nu=1/d$.  However, as noted in
Refs.~\cite{PF-83,FP-85}, finite-size effects strongly depend on the
geometry, differing significantly in cubic $L^d$ and anisotropic
geometries. For instance, in $L^{d-1}\times M$ geometries with $M\gg
L$, FSS in Ising systems is characterized by exponential laws in
$L$~\cite{PF-83}.  Recent studies of quantum FOTs have also reported a
significant dependence on the boundary
conditions~\cite{CNPV-14,CPV-15}.

In this paper, we study the static and dynamic behavior at thermal
FOTs in finite-size systems with boundary conditions favoring one of
the two phases, in particular the disordered phase, such as open
boundary conditions (OBC).  We consider the two-dimensional (2D) Potts
model, which is a standard theoretical laboratory to understand issues
concerning the statistical behavior at a thermal FOT.  We show, in the
presence of OBC, the emergence of an {\em anomalous} equilibrium FSS
(EFSS), which is characterized by scaling laws that differ from those
that apply in the coexistence region with PBC.  Boundary conditions
have also a crucial influence when considering dynamic phenomena, for
instance, when one considers a slow heating of the system, starting in
the ordered phase and moving across the FOT.  We show that disordered
boundary conditions give rise to an off-equilibrium FSS (OFSS)
characterized by a time scale increasing as a power of the size
$\ell$, i.e. $\tau(\ell)\sim\ell^2$.  We argue that these static and
dynamic features are shared among generic thermal FOTs, when the
boundary conditions favor one of the two phases.

Before presenting our results, we discuss the phenomenological
relevance of our study.  While PBC are not usually appropriate to
describe realistic situations, OBC are appropriate whenever the
disordered phase is somehow favored by the boundaries.  Such a
situation arises in many physical contexts.  One notable example is
provided by heavy-ion collision experiments, probing the phase diagram
of hadronic matter.  At finite temperature ($T$), theoretical
arguments~\cite{RW-00} predict the existence of a high-$T$ quark-gluon
phase and of a low-$T$ hadronic phase, separated by a FOT line at
finite quark chemical potential, ending at an Ising-like transition
point.  Heavy-ion collision experiments are trying to find evidence
for such FOT line~\cite{qgp-ref,BGSG-99,HHMN-13}; see, in particular,
the planned activities discussed in Ref.~\cite{Petersen-17} and
references therein.  One major problem in identifying its signature is
the presence of space-time inhomogeneities in the quark-gluon plasma
generated in the collisions. The plasma is expected to be confined in
a small region with a size of a few femtometers (fm), and to hadronize
within a time interval of a few ${\rm fm}/c$.  In the appropriate
region of the phase diagram, as the plasma cools down, the system is
expected to cross the FOT line.  However, as the size and time scales
are finite, there must be a substantial rounding of the FOT
singularities, which may conceal its presence. Therefore, for a
correct interpretation of the experimental results, it is important to
understand the effects of space-time inhomogeneities at FOTs.  Since,
for the hadronic transition, the high-$T$ and low-$T$ phases are the
ordered and the disordered ones, respectively (such correspondence is
the opposite of the usual one)~\cite{RW-00}, the quark-gluon plasma
dynamics corresponds to that of a finite-size statistical system, such
as the Potts model, that is slowly heated across a FOT.  Note also
that hadron matter surrounds the quark-gluon plasma, hence the
appropriate boundary conditions in the corresponding statistical
system must favor the disordered phase, as is the case for the OBC.
The dynamics across the FOT of a finite-size system with OBC should
therefore capture some of the important features of the evolution of a
confined quark-gluon plasma cooled down across the FOT toward the
hadronic low-$T$ phase. This may lead to a better understanding of the
signatures of the FOT line in heavy-ion
collisions~\cite{footnoteew-SR-12}.

The 2D $q$-state Potts model provides a theoretical laboratory to
study thermal FOTs. It is defined by
\begin{equation}
Z=\sum_{\{s_{\bf x}\}} e^{-\beta H},\qquad
H =  - \sum_{\langle {\bf x}{\bf y}\rangle} \delta(s_{{\bf x}}, s_{ {\bf y}}), 
\label{potts}
\end{equation}
where $\beta\equiv 1/T$, the sum is over the nearest-neighbor sites of
an $L\times L$ lattice, $s_{\bm x}$ are integer variables $1\le
s_{{\bm x}} \le q$, $\delta(a,b)=1$ if $a=b$ and zero otherwise.  It
undergoes a phase transition~\cite{Baxter-book,Wu-82} at
$\beta_c\equiv1/T_c= \ln(1+\sqrt{q})$, which is of first order for
$q>4$.  At $T_c$ the infinite-volume energy density $E \equiv \langle
H \rangle/L^2$ is discontinuous. We define
\begin{eqnarray}
E_r \equiv  
\Delta_e^{-1}\,(E -E_c^-),\qquad \Delta_e \equiv E_c^+ -
E_c^-,
\label{ener}
\end{eqnarray}
where $E_c^\pm \equiv E(T_c^\pm)$~\cite{results-q20} , so that
$E_r=0,1$ for $T\to T_c^-$ and $T\to T_c^+$, respectively.  The
magnetization $M_k$
\begin{eqnarray}
M_k = \langle \mu_k \rangle, \qquad  
\quad \mu_k  \equiv 
{1\over V}\sum_{\bf x} {q \delta(s_{\bf x},k) - 1\over q-1},
\label{madef} 
\end{eqnarray}
is also discontinuous, i.e. ${\rm lim}_{T\to T_c^-} M_k =
m_0>0$~\cite{results-q20}.  

One can define an EFSS
~\cite{NN-75,FB-82,PF-83,FP-85,CLB-86,BK-90,VRSB-93,CNPV-14} also at
FOTs, with scaling laws that are analogous to those holding at
continuous transitions~\cite{Barber-83,Privman-90}.  This issue has
been mostly investigated for cubic $L^d$ systems with PBC, finding
that the finite-size coexistence region shows a scaling behavior in
terms of the variable $u\propto \delta\,L^d$ with $\delta \equiv 1 -
\beta/\beta_c$. For instance, the energy density scales as $E_r(T,L)
\approx {\cal E}(u)$. For the 2D Potts model ${\cal E}(u)=(1 + q
e^{-u})^{-1}$~\cite{footnotepbcefss}. Systems with boundary conditions
favoring the high-$T$ phase, such as the OBC, show a more complex
behavior, due to surface effects, which give rise to a shift
$\delta^*(L)\sim 1/L$ of the transition temperature. Correspondingly
the scaling variable to describe the coexistence regime is
$u\propto[\delta - \delta^*(L)] L^2$~\cite{PR-90}, whose check
requires a precise determination of $\delta^*(L)$.  However, as we
shall see, other interesting EFSS properties emerge.

\begin{figure}[tbp]
\includegraphics*[scale=\graphicscale]{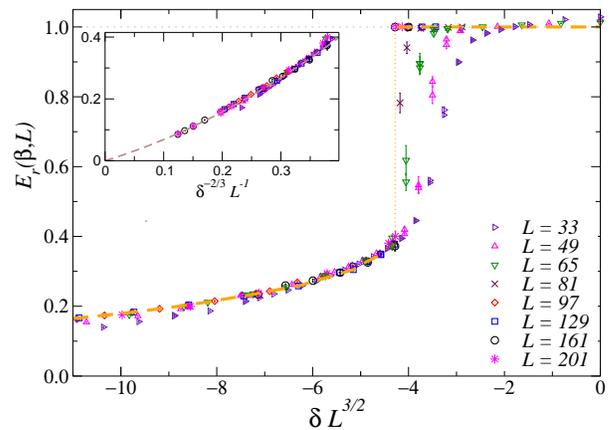}
\caption{ Equilibrium data for the renormalized energy density $E_r$,
  cf. Eq.~(\ref{ener}), for systems with OBC around the FOT point,
  versus $w\equiv \delta\,L^{\varepsilon}$ with $\varepsilon=3/2$.  By
  increasing the size of the system, the data appear to approach a
  nontrivial FSS curve (dashed line).  The inset shows the data of the
  low-$T$ region, versus $w^{-1/\varepsilon}=\delta^{-1/\varepsilon}
  L^{-1}$; the dashed line is a fit to ${\cal E}_e(w)\approx a_1
  |w|^{-1/\varepsilon} + a_2 |w|^{-2/\varepsilon}$ ($a_1\approx 0.6$
  and $a_2\approx 0.9$).  }
\label{equi}
\end{figure}

\begin{figure}[tbp]
\begin{tabular}{lr}
\includegraphics*[width=4.4truecm,angle=0]{twopeaks.eps} &
\includegraphics*[width=3.2truecm,origin=c,angle=-90]{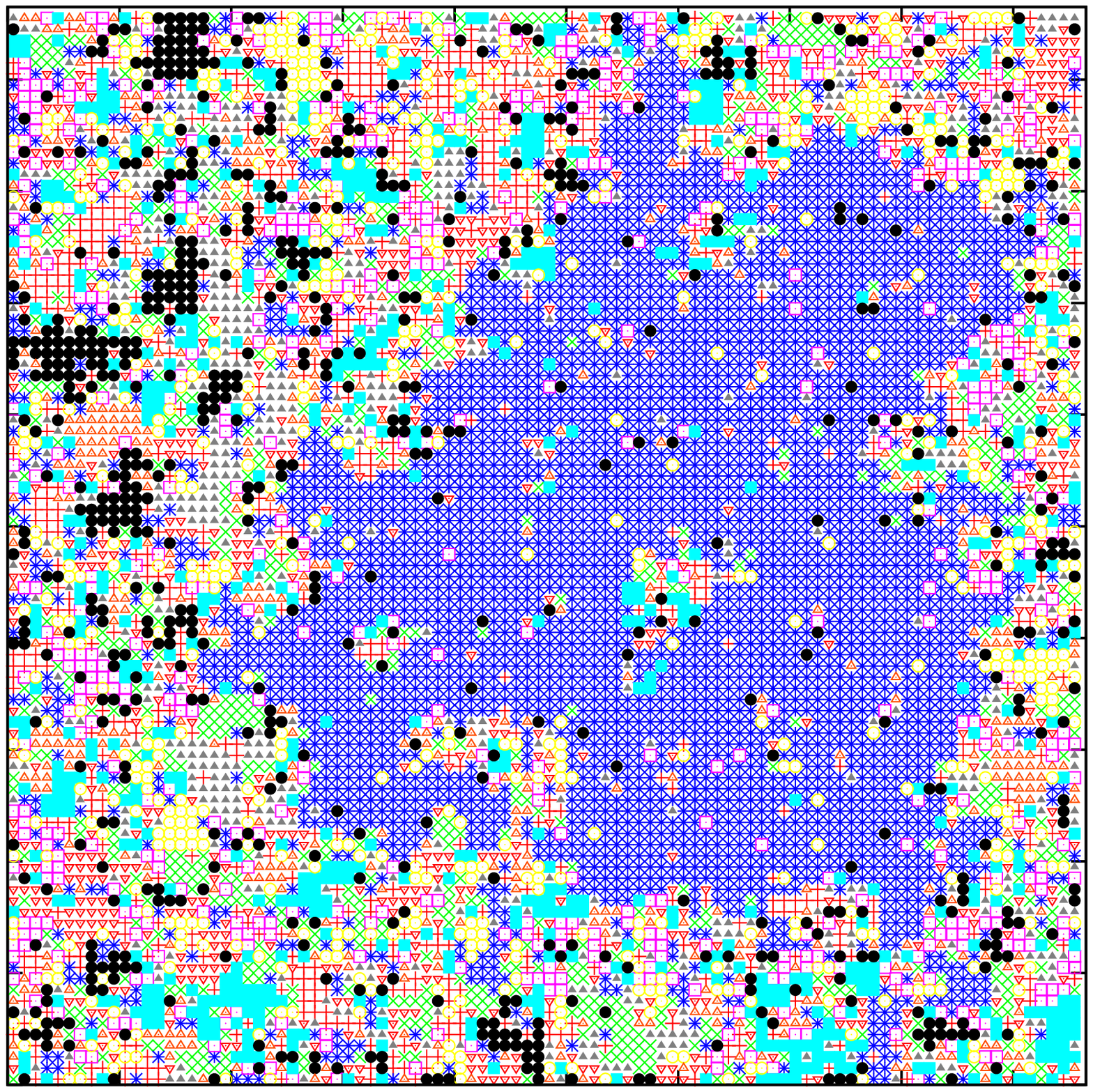}  \\
\end{tabular}
\caption{(Left) Distribution of $E_r$ for various values of $L$ at 
  $w = w^*$, which is where the specific heat has a maximum (this is
  very close to the value where the areas below the two peaks are
  equal).  Here $q=10$. (Right) A typical ``low-temperature" configuration
  with $L=192$ at $w = w^*$: the same color and symbol corresponds to
  the same value of the spin.  }
\label{fig:coex}
\end{figure}

To investigate these issues, we have performed simulations with $q=20$
(up to $L=200$) and $q=10$ (up to $L=512$) using OBC~\cite{heatbath}.
Quite surprisingly, see Fig.~\ref{equi}, the results appear to scale
as
\begin{equation}
E_r(\delta,L) \approx {\cal E}_e(w), \qquad w=\delta\,L^\varepsilon,
\label{efss}
\end{equation}
where $\varepsilon \approx 3/2$ (the optimal collapse of the data
suggest it with an accuracy on $\varepsilon$ of a few per cent).
This nontrivial scaling is observed for $w<0$ up to
an $L$-dependent value $w^*(L)$.  Since $E_r = 0$ in the $L\to\infty$
limit for $\delta = 0^-$, we expect ${\cal E}_e(w\to -\infty)\to 0$.
Moreover, since for OBC the $L\to\infty$ limit at fixed $T$ is
approached with $1/L$ corrections, we expect ${\cal E}_e(w)\approx
|w|^{-1/\varepsilon}$ for $w\ll -1$.  This is confirmed by the data
for $w\lesssim w^*$, see the inset of Fig.~\ref{equi}.  For $w >
w^*(L)$, the scaling behavior is trivial as ${\cal E}_e(w) = 1$, the
high-$T$ value, with $1/L$ corrections.  Analogous results are
obtained for other observables and
for $q=10$~\cite{footnoteconv}.  For $w=w^*(L)$ the energy has a
two-peak structure, see Fig.~\ref{fig:coex}, and the probabilities of
the corresponding configurations are approximately equal. The value
$w^*(L)$ is therefore related to the shift $\delta^*(L)$ of the
transition~\cite{footnotewstar}.  Note that the scaling in terms of
$w$ is not appropriate to describe coexistence, but only the low-$T$
region $w < w^*$.  Indeed, when expressed in terms of $w$, the EFSS
functions are singular for $w = w^*$, and the scaling is trivial for
$w>w^*$.

To interpret the scaling in terms of $w$, we note that, for $w <
w^*(L)$, typical configurations are characterized by a central ordered
region surrounded by a disordered ring of volume $V_+$, stabilized by
the boundary conditions, see the right panel of Fig.~\ref{fig:coex}.
Thus, the size of the disordered region can be related to $E_r$, by
$E_r\approx V_+/L^2$. At fixed $w$, the energy $E_r$ is fixed,
implying that the scaling limit we consider is appropriate to describe
low-$T$ configurations characterized by the simultaneous presence of
disordered and ordered regions.  One should not confuse this mixed
regime with thermodynamic coexistence: for $w < w^*(L)$ the free
energy has only one minimum (thermodynamically it represents a low-$T$
state), except in a small interval of $w$ around $w^*$, which shrinks
as $L$ increases.  Therefore, the scaling (\ref{efss}) is appropriate
to describe the low-$T$ side of the transition, in an interval $\delta
\sim L^{-\varepsilon}$ where a significant part of the volume is
disordered. Note that, since $|w^*(L)|$ increases with
$L$~\cite{footnotewstar}, the region of the scaling (\ref{efss}) is
restricted to larger and larger values of $|w|$ with increasing $L$.
Since boundary effects cannot extend beyond a length scale $\xi$,
$V_+$ is limited by $L\xi$, and $E_r(w^*)$ decreases as $L$ increases.
Nevertheless, this regime turns out to be relevant for the dynamic
behavior across the FOT, see below.

We have checked that analogous EFSS behaviors are observed for
different geometries and boundary conditions, for example when we
consider OBC in the first direction and PBC in the second one, and in
the case of fixed boundary conditions in which all spins are equal on
the boundary, thus favoring the ordered phase.

We now discuss the dynamical behavior of the system, assuming that it
is slowly heated across the FOT, starting from the low-$T$ phase.  We
consider the linear protocol
\begin{equation}
\delta(t)  \equiv {1 - \beta(t)/\beta_c} = t/t_s
\label{betat}
\end{equation}
where $t\in [t_i<0,t_f>0]$ is a time variable varying from a negative
to a positive value, and $t_s$ is the time scale of the process.  The
value $t=0$ corresponds to $\beta(t)=\beta_c$.  The dynamics starts
from an ordered configuration with $M_1 \approx m_0$,
cf.~Eq.~(\ref{madef}), at $\beta_i= \beta_c [1 -
  \delta(t_i)]>\beta_c$, so that $\delta(t_i) <
\delta^*(L)$~\cite{footnotedepic}.  Then, the system evolves under a
heat-bath MC dynamics~\cite{heatbath}, which corresponds to a purely
relaxational dynamics~\cite{HH-77}. The time unit is a heat-bath sweep
of the whole lattice. The temperature is changed according to
Eq.~(\ref{betat}) every sweep, incrementing $t$ by one.  We repeat
this procedure several times averaging the observables, such as the
energy density $E(t,t_s,L) = \langle H \rangle_t/V$ and magnetization
$M_1(t,t_s,L) = \langle \mu_1 \rangle_t$, over the ensemble of
configurations obtained by the off-equilibrium protocol at time $t$.

In systems with OBC the transition from the low-$T$ to the high-$T$
phase occurs through a mixed-phase regime, which is related to the
dynamics of a closed domain wall, with a relatively small (negligible
in the large-$L$ limit) thickness, that spatially separates the two
coexisting phases, see Fig.~\ref{fig:coex}.  For $w \ll w^*$, such a
domain wall is localized at the boundaries, thus $E_r\approx 0$.  With
increasing $w$, it moves toward the center of the lattice, until the
whole volume is in the high-$T$ phase, thus $E_r\approx 1$. The
transition occurs through a free-energy barrier, hence we expect the
system to be out of equilibrium as one moves from one phase to the
other.

\begin{figure}[tbp]
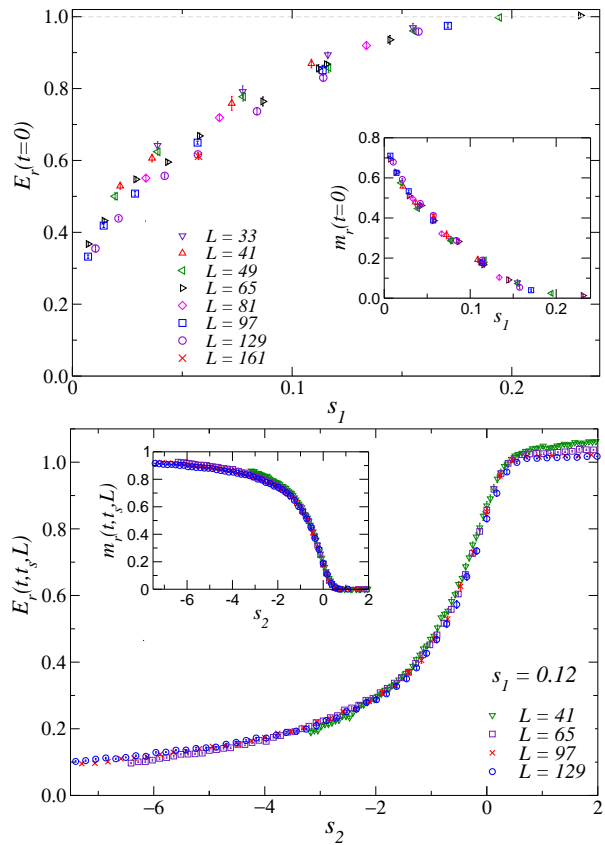

\includegraphics*[scale=\graphicscale]{ermrt0.eps}
\includegraphics*[scale=\graphicscale]{emtau.eps}
\caption{Plots of $E_r$ and $m_r$ (insets) at $t=0$ (top) and fixed
  $s_1=0.12$ (bottom).  They support the OFSS predictions.  }
\label{offdata}
\end{figure}

To identify a scaling regime that describes the dynamic behavior, we
must specify appropriate scaling variables. First, we wish to recover
the EFSS defined in Eq.~(\ref{efss}) in the appropriate limit, see
below. Thus, the corresponding dynamic scaling limit should be defined
at fixed $r_1 = \delta(t) L^\varepsilon = t L^\varepsilon/t_s$. The
second scaling variable must be $t/\tau(L)$, where $\tau(L)$ is the
time scale of the dynamics. The identification of $\tau(L)$ is
strictly dependent on the choice of the EFSS variable.  At fixed
$r_1$, the transition region in which the free energy shows a
double-peak structure shrinks as $L\to \infty$. Therefore, the
mixed-disordered coexistence region at $\delta^*$ is not relevant for
the scaling described here. The time interval $\Delta t$ that the
system spends in this region is vanishingly small compared to the
relevant time scales: $\Delta t \, L^\varepsilon/t_s \to 0$ in the
scaling limit. The transition from the ordered to the disordered phase
occurs for $r_1 > w^*$, i.e., when the free energy has only the
high-$T$ minimum.  Under these conditions, we expect $\tau(L) \sim
L^z$ with $z = 2$, which is the expected behavior of the time scale
for the shrinking of an ordered domain surrounded by the more stable
disordered phase under a purely relaxational dynamics~\cite{Bray-94}.
Therefore, the relevant scaling variables are $r_1
=(t/t_s)\,L^\varepsilon$ and $r_2 = t/L^z$, or their combinations
\begin{eqnarray}
s_1 
= {t_s/L^{\varepsilon + z}},\qquad
s_2 
={t/t_s^{z/(\varepsilon+z)}}.
\label{s2def}
\end{eqnarray}
OFSS develops in the limit $t,t_s,L\to\infty$ keeping $s_1$ and $s_2$
fixed.  For example, we expect
\begin{eqnarray}
&&E_r(t,t_s,L) \equiv {E(t,t_s,L) -E_c^-\over E_c^+-E_c^-} 
\approx {\cal E}_{o}(s_1,s_2). 
\label{erscat}
\end{eqnarray}
Analogously $m_r \equiv M_1(t,t_s,L)/m_0 \approx {\cal M}_o(s_1,s_2)
$.  The EFSS of Eq.~(\ref{efss}) must be recovered for large values of
$s_1$ keeping $r_1$ fixed, where ${\cal E}_o(s_1,s_1^{1 +
  z/\varepsilon} r_1) \approx {\cal E}_e(r_1)$.  The MC results fully
support the above OFSS.  In particular, Fig.~\ref{offdata} shows that
the data at $t=0$, thus $s_2=0$, approach a function of
$s_1=t_s/L^{7/2}$, and the data at fixed $s_1\approx 0.12$ approach a
function of $s_2=t/t_s^{4/7}$ (analogous results are obtained for
other values of $s_1$).

It is important to stress the difference between the types of scaling
reported above and those discussed in the literature. For instance,
for PBC the relevant time scale scales exponentially~\cite{PV-17},
$\tau(L)\sim \exp(\sigma L)$, as it is related to the tunneling time
between the two phases. This occurs by means of the generation of
strip-like configurations with two interfaces, whose probability is
suppressed by a factor $\exp(-\sigma L)$, where $\sigma = 2 \beta_c
\kappa$ and $\kappa$ is the interface tension~\cite{results-q20}.  The
different behavior observed here has been made manifest by the
particular choice of the static scaling variable, which allows us to
focus on an emergent mixed regime in systems with OBC.  If we were
considering an EFSS appropriate to describe the coexistence region, we
would also obtain an exponential behavior $\tau(L) \sim e^{\alpha L}$,
where $\alpha$ is related to the height of the free-energy barrier
present at coexistence.
These different scalings can be rephrased by considering $t_s$. In the
OBC case, two different regimes appear. There is an intermediate
regime in which $t_s$ scales as $L^{\varepsilon+z} = L^{7/2}$ and we
observe scaling in terms of $s_1$ and $s_2$. On the other hand, for
$t_s$ much larger, $t_s \sim L^p e^{\alpha L}$, one should observe a
different scaling appropriate to  the dynamics in the
coexistence region.

Although the numerical data strongly support the existence of the
anomalous EFSS, with corresponding OFSS, some open questions still
remain, calling for additional studies.  In particular, we are not
able to predict the exponent $\varepsilon$, which is likely related to
the dynamics of the domain wall separating the ordered and disordered
phase regions, and its dependence on the spatial dimension.

Our results show that the behavior of mixed phases at FOTs strongly
depend on the boundary conditions.  In particular, in systems with
boundary conditions that favor one of the two phases there exists a
scaling regime characterized by the physical coexistence of both
phases: the thermodynamically favored phase coexists with the
unfavored one, which is stabilized by the boundary conditions. In this
scaling regime, in which $\delta L^\varepsilon$ is the appropriate
scaling variable, if one slowly heats the system across the
transition, one may define a dynamical scaling with a characteristic
time $\tau(L)$ that scales as $L^2$.  We expect the main features
observed in the Potts model with OBC to be generally shared by systems
at thermal FOTs when boundary conditions favor one of the two phases.
Moreover, we expect that the main properties of the off-equilibrium
behavior across the FOT also hold for other types of dynamics, for
example in the presence of conservation laws.  We note that power-law
OFSS are generally observed at continuous
transitions~\cite{GZHF-10,CEGS-12,PRV-18}.  Our study shows that the
main distinguishing feature of the OFSS across a FOT is related to the
existence of a mixed regime, where the two phases are spatially
separated.

Unlike PBC, boundary conditions favoring one of the two phases are of
experimental interest.  Their EFSS and OFSS may be exploited to gain
evidence of a FOT in experiments with finite-size systems in
off-equilibrium conditions.  For example, they are relevant for
heavy-ion experiments~\cite{Petersen-17} searching for evidences of
FOTs in the hadron-matter phase diagram~\cite{RW-00}.  According to
our results, when the quark-gluon plasma cools down across the FOT
line, a scaling behavior can be observed on time scales $\tau \sim
\ell^2$, where $\ell$ is the size of quark-gluon plasma, expected to
be a few fm.  Note that, since the cooling is relatively
fast---typical times should be of the order of a few fm/$c$ ---this is
probably the only scaling behavior that can be observed in practice.
As already noted, power-law OFSS behaviors also characterize
continuous transitions; however, the key feature of the crossing of a
FOT line should be that it occurs through a mixed regime, where the
hadronic and quark-gluon phases are spatially separated.

We finally mention that {\em anomalous} EFSS, and corresponding OFSS,
are expected to also arise at quantum FOT of many-body systems whose
boundary conditions favor one of the two phases.  The recent great
progress in the control of isolated small quantum
systems~\cite{Bloch-08, GAN-14} may allow their experimental
investigations.


\begin{thebibliography}{99}
\bibitem{footnotepbcefss} This is obtained by expressing the partition
  function of mixed states around $T_c$ as a sum of two exponential
  functions related to the free energies of the pure high-$T$ and
  low-$T$ phases~\cite{CLB-86,BK-90,PV-17}, neglecting exponentially
  suppressed contributions related to the interface.  The scaling
  variable turns out to be $u = \Delta_e \beta_c \delta L^2$.
\bibitem{heatbath} We use a standard heat-bath algorithm, which updates a single
  site variable by generating a new spin $s_{\bm x}'$ at ${\bm x}$
  with probability $\sim e^{-H(s_{\bm x}')/T}$ independent of the old
  spin.
\bibitem{footnotewstar} 
We expect a
  surface-induced shift of the coexistence point~\cite{PR-90}: in finite volume the
  pseudocritical point should correspond to $\delta^* \sim 1/L$. Hence
  $w^*(L)$ should scale as $w_0 L^{1/2}$. For $q=20$ such a shift
  cannot be observed in our range of values of $L$; apparently, the
  proportionality constant $w_0$ is very small. For $q=10$, we observe
  an $L$-dependence of $w^*$, although we are not able to confirm the
  expected $L^{1/2}$ behavior.
\bibitem{footnotedepic} The resulting asymptotic OFSS around
  $\delta(t)=0$ does not depend on the initial finite value of $\delta_i<0$,
  because it develops in a narrow range around $\delta(t)\approx 0$.
  It does not depend on the final value $\delta_f>0$ either.
\end{thebibliography}
\end{document}